\documentclass[iop]{emulateapj}

\newcommand{\be}{\begin{eqnarray}}
\newcommand{\ee}{\end{eqnarray}}
\newcommand{\lp}{\left(}
\newcommand{\rp}{\right)}

\newcommand{\E}[1]{\times10^{#1}}
\newcommand{\smpy}{ \ M_\odot \ {\rm yr}^{-1}}
\newcommand{\msol}{ \ M_\odot }
\newcommand{\commentOut}[1]{}
\newcommand{\bi}{\begin{itemize}}
\newcommand{\ei}{\end{itemize}}
\newcommand{\cgsd}{ {\rm \ g \ cm^{-3}}}
\newcommand{\cgsv}{ {\rm \ cm \ s^{-1}}}
\newcommand{\dd}[2]{ \frac{ \partial #1 }{ \partial #2 } }
\newcommand{\grad}{\vec{\nabla}}

\shorttitle{DOUBLE WD MERGERS}
\shortauthors{SHEN ET AL.}
\slugcomment{Submitted to The Astrophysical Journal}

\begin{document}


\title{The Long-Term Evolution of Double White Dwarf Mergers}

\author{Ken J. Shen\altaffilmark{1,2,3}, Lars Bildsten\altaffilmark{4}, Daniel Kasen\altaffilmark{1,5}, \& Eliot Quataert\altaffilmark{2}}
\altaffiltext{1}{Lawrence Berkeley National Laboratory, Berkeley, CA 94720, USA; kenshen@astro.berkeley.edu.}
\altaffiltext{2}{Department of Astronomy and Theoretical Astrophysics Center, University of California, Berkeley, CA 94720, USA.}
\altaffiltext{3}{Einstein Fellow.}
\altaffiltext{4}{Kavli Institute for Theoretical Physics, Kohn Hall, University of California, Santa Barbara, CA 93106, USA.}
\altaffiltext{5}{Department of Physics, University of California, Berkeley, CA 94720, USA.}


\begin{abstract}

In this paper, we present a model for the long-term evolution of the merger of two unequal mass C/O white dwarfs (WDs).  After the dynamical phase of the merger, magnetic stresses rapidly redistribute angular momentum, leading to nearly solid-body rotation on a viscous timescale of $10^{-4}-1$ yr, long before significant cooling can occur.  Due to heating during the dynamical and viscous phases, the less massive WD is transformed into a hot, slowly rotating, and radially extended envelope supported by thermal pressure.

Following the viscous phase of evolution, the maximum temperature near the envelope base may already be high enough to begin off-center convective carbon-burning.  If not, Kelvin-Helmholtz contraction of the inner region of the envelope on a thermal timescale of $10^3- 10^4$ yr compresses the base of the envelope, again yielding off-center burning.  As a result, the long-term evolution of the merger remnant is similar to that seen in previous calculations: the burning shell diffuses inwards over $\sim 10^4$ yr, eventually yielding a high-mass O/Ne WD or a collapse to a neutron star, rather than a Type Ia supernova.  During the cooling and shell-burning phases, the merger remnant radiates near the Eddington limit.  Given the double WD merger rate of a few per 1000 yr, a few dozen of these $\sim 10^{38}$ erg s$^{-1}$ sources should exist in a Milky Way-type galaxy.

While the end result is similar to that of previous studies, the physical picture and the dynamical state of the matter in our model differ from previous work.  Furthermore, substantial remaining uncertainties related to the convective structure near the photosphere and mass loss during the thermal evolution may significantly affect our conclusions.  Thus, future work within the context of the physical model presented here is required to better address the eventual fate of double WD mergers, including those for which one or both of the components is a He WD.

\end{abstract}



\section{Introduction}

Type Ia supernovae (SNe Ia; \citealt{hn00}) are well known for their role in determining the accelerating expansion of the Universe \citep{ries98,perl99}.  While the generally accepted model of a thermonuclear runaway in a C/O white dwarf (WD) explains many of the defining characteristics of SNe Ia, substantial uncertainty remains as to the progenitor system.  Three progenitor scenarios are currently considered: (1) the single degenerate scenario (SDS; \citealt{wi73,nomo82a}), in which a WD accretes and stably burns H-rich material from a non-degenerate main sequence, sub-giant, or red giant donor; (2) the double degenerate scenario, in which two C/O WDs undergo a dynamical merger \citep{it84,webb84}; and (3) the double detonation scenario \citep{livn90,lg90,lg91,ww94,la95}, in which a detonation in an overlying He-burning shell sends a converging shock wave into the C/O core that initiates a subsequent detonation near the center.

In the SDS and double degenerate scenario, convective C-burning is ignited by strongly-screened or pycnonuclear reactions induced by very high densities $\sim 10^{10} \cgsd$ in the center of a near-Chandrasekhar mass WD.  After $300-1000$ yr \citep{wwk04,pc08}, the burning in this convective region becomes violent enough to yield a deflagration and/or detonation, incinerating the WD.  In the double detonation scenario, no core convective burning takes place: the shock from a He shell detonation initiates a secondary detonation in the C/O core that disrupts the WD.  While these scenarios cover a wide range of possibilities, none can adequately explain specific features of SNe Ia and their observed rate.

The SDS relies on the stable burning of accreted hydrogen in order to grow the WD.  However, it has been known for decades \citep{sien80,fuji82b,pacz83} and recently confirmed \citep{ylv04,sb07,nomo07} that the range of accretion rates that allow for stable H-burning is very narrow.  Furthermore, the accumulated He layer will undergo nova-like flashes that likely eject some of the accreted material.  As a result, population synthesis calculations \citep{bran95,yl98,yl00,rbf09} predict SN Ia rates from the SDS channel that are $\sim 10$ times lower than the observed rate \citep{mann05,sb05,dild08}.  Additionally, observational limits on the progenitor systems \citep{dkp06,dist10a,gb10,li11} and on the SN Ia ejecta's interaction with a donor \citep{mbf00,kase10,hayd10,bian11,nuge11,bloo12} severely constrain the red giant donor channel's contribution to the SN Ia population.

The double detonation scenario has received more attention recently due to the prediction of He detonations in pre-AM CVn systems \citep{bild07,fhr07,fink10,sb09b,shen10,sim10,wk11}.  However, while many features of this scenario are attractive, it is not clear how the He-detonation products on the outside of the core will skew observations away from SN Ia-like events.  Even a small amount of surface iron-group elements created during He-burning, especially Ti and Cr, can significantly impact the colors of the SN at peak light.  The UV line blanketing due to these elements causes the predicted spectra near peak light to be much redder than observed \citep{krom10,sim11}.\footnote{Conversely, previous studies found the colors to be too blue due to $^{56}$Ni decays near the photosphere \citep{hk96,nuge97}.}  Furthermore, the He detonation will eject radioactive isotopes at high velocities, a feature that has never been observed in SNe Ia.


\subsection{An updated model for double white dwarf mergers}

The double degenerate scenario involves the dynamical coalescence of two C/O WDs.  The predicted merger rate and delay time distribution \citep{bran95,nele01a,rbf09} and observations of progenitor systems (albeit with large error bars; \citealt{nele05b,geie10a,geie10b,tovm10,rodr10}) are promisingly consistent with the observed SN Ia rate and delay time distribution \citep{msg10,maoz11,grau11}.  However, previous theoretical studies have found that the merger of two WDs instead triggers an accretion induced collapse, yielding a faint explosion and a neutron star remnant instead of a SN Ia \citep{ni85,ksn87,sn04}.

These studies assumed that the less massive WD is disrupted into a Keplerian disk from which the more massive WD accretes at a constant rate near the gravitational Eddington limit of $10^{-5} \smpy$ for $\sim 10^5$ yr.  In contrast, smoothed particle hydrodynamical (SPH) studies of the merging process \citep{benz90,rs95,scm97,ggi04,lore05,lig09,ypr07,dan11,rask11} find that the tidally disrupted WD is instead converted into material that is supported against gravity by a combination of thermal pressure from shock heating in regions close to the more massive WD and centrifugal support in the optically thick outer regions.  Until the work of \cite{ypr07} and \cite{vcj10}, these important features of the merged configuration were largely neglected in studies of the long-term evolution of merger remnants.

One conclusion reached from consideration of this more detailed merger end-state is that the viscous evolution of the differentially rotating layers must be taken into account before the secular thermal evolution of the merger remnant can be calculated.  In particular, if magnetically-governed processes such as the magnetorotational instability \citep{bh91,hb91,balb03} and the Tayler-Spruit dynamo \citep{tayl73,spru99,spru02} operate efficiently, angular momentum will be redistributed throughout the remnant much more rapidly than heat can be radiated away.  \emph{Thus, the envelope consisting of the tidally disrupted WD will transfer most of its angular momentum to a small amount of mass in the outer layers and approach a viscously-heated shear-free configuration long before cooling significantly.}

In this paper, we present an updated physical model for the evolution of unequal mass double WD mergers.  Dynamical and viscous processes, acting on timescales much shorter than the thermal timescale, convert the majority of the less massive WD into a hot radially extended envelope surrounding the more massive WD.  Little mass remains centrifugally-supported following the viscous evolution.  While the entire remnant has a mass essentially equal to the total mass of the initial binary and may be larger than the Chandrasekhar mass, the hot envelope is extended and does not exert considerable pressure on the degenerate core.  

Substantial heat is generated during the dynamical and viscous phases of evolution and may lead to convective C-burning at the interface between the hot envelope and degenerate core during or directly after the viscous evolution.  If instead the maximum temperature is too low for immediate convective C-burning prior to significant thermal transport, the inner part of the envelope will undergo a Kelvin-Helmholtz contraction on a thermal timescale of $10^3- 10^4$ yr, which is set by the Eddington luminosity of the giant envelope.  The hot non-rotating envelope radiates near the Eddington limit, cooling and compressing the base of the envelope relatively rapidly, just as the base of the accreted envelope is compressed in studies assuming disk accretion such as \cite{ni85}.  This contraction proceeds on a shorter timescale than the inward thermal diffusion timescale into the degenerate core of $\sim 10^6$ yr, so that a temperature inversion is formed and a convective C-burning shell is born well away from the center.

Thus, it appears unavoidable that double C/O WD mergers ignite C-burning well away from the merger remnant's center.  This off-center ignition allows the convective layer to expand, and the necessary conditions for a violent deflagration and/or detonation are not reached.  Under the assumption of heat transport by thermal conduction, the burning shell moves slowly inwards at a speed of $\sim 0.01 \cgsv$, reaches the center in $\sim 10^{4}$ yr, and eventually converts the previously C/O WD into an O/Ne WD \citep{twt94,sn98}.  If the remnant mass is larger than the Chandrasekhar mass, envelope material can continue to contract and compress the O/Ne WD until it collapses into a neutron star, accompanied by a weak explosion.  While such events will not yield SNe Ia, they may give rise to fast transients and contribute to $r$-process enrichment, and are certainly of astrophysical interest \citep{dess06,mpq09a,mpq09b,darb10}.

Although this chain of events represents an updated description of double WD merger evolution, the final outcome is similar to that found by previous studies assuming disk accretion.  The onset of convective C-burning in a shell following the viscous phase of evolution, either prior to or after a period of thermal redistribution, seems inevitable.  However, the physical picture leading to this evolution is markedly different from previous work.  In our model, \emph{there is no phase of accretion from a Keplerian disk onto a degenerate object; instead, the long-term thermal evolution of the merger remnant is determined by the conditions at the base of the hot non-rotating envelope and the rate at which it redistributes heat and cools.}

In this paper, we describe our updated model for the evolution of a double C/O WD merger, focusing separately on the dynamical (\S \ref{sec:dynevol}), viscous (\S \ref{sec:viscevol}), and thermal (\S \ref{sec:thermevol}) phases.  In \S \ref{sec:comp}, we compare to previous work.  We conclude in \S \ref{sec:conc} and speculate on future work and potential observables.


\section{Dynamical evolution: $10^2-10^3$ \lowercase{s}}
\label{sec:dynevol}

Close binaries consisting of two C/O WDs are formed via two common envelope phases \citep{it84}: one while the initially more massive star is on its asymptotic giant branch (AGB), and another during the second star's AGB or at the tip of its red giant branch.  In the latter case, the second star lives for $\sim 100$ Myr as a core He-burning sdO or sdB star before becoming a C/O WD \citep{hebe09}.  The orbital separations of double WD binaries shrink via gravitational wave radiation, bringing WD binaries with initial orbital periods $ \lesssim 10$ hr into mass transfer contact within a Hubble time.  These binaries are being discovered in growing numbers by studies such as the SPY and the SDSS-based SWARMS surveys \citep{york00,nele05b,bade09,mull09,geie10a}.

Upon contact, the less massive WD overflows its Roche lobe and begins transferring matter onto the more massive WD.  The dynamical stability of this mass transfer depends on the mass ratio and on how efficiently angular momentum is returned to the orbit during mass transfer \citep{mns04}.  For typical assumptions, mass transfer between two C/O WDs is dynamically unstable, even for the lowest expected mass ratio of $0.5/1.2$.

The unstable mass transfer is driven by the less massive WD's hydrodynamical response to mass loss, and thus the dynamical evolution is characterized by that WD's sound crossing time or, nearly equivalently, the binary's orbital period: $30-300$ s.  The subsequent evolution of the resulting merger remnant is dominated by the transport of angular momentum and entropy.  Simple estimates of the importance of magnetic stresses lead to viscous evolution within $10^4-10^8$ s.  This timescale is much longer than the dynamical timescale, but much shorter than the remnant's thermal timescale, $10^3- 10^4$ yr.  Thus, we separate discussion of the merger and the subsequent evolution of the remnant into three distinct phases based on this hierarchy of timescales.

The dynamical merging process has been studied in detail \citep{benz90,rs95,ggi04,lig09,dan11,rask11}, so we only emphasize a few results.  Once mass transfer begins, material leaves the less massive WD's Roche nozzle and directly impacts the surface of the more massive WD \citep{webb84,mns04}.  The resulting shock efficiently converts the gravitational and kinetic energy into thermal energy, so that this material is nearly virialized.  As tidal disruption progresses in earnest, the less massive WD becomes unbound from itself.  Because much of this material has not been shocked, its entropy stays relatively low, and it remains supported against gravity by centrifugal forces as it orbits the more massive WD.

\begin{figure}
	\plotone{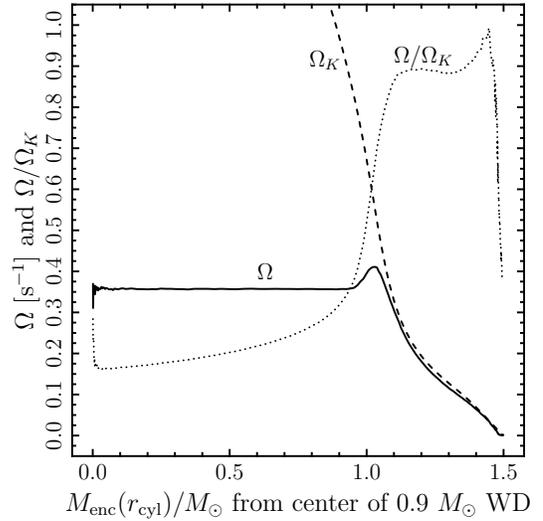}
	\caption{Angular velocity for material within $10^8$ cm of the equatorial midplane (\emph{solid line}), Keplerian angular velocity (\emph{dashed line}), and their ratio (\emph{dotted line}) vs. cylindrically enclosed mass from the center of the $0.9 \msol $ WD.  Data are taken at the end of an SPH simulation of a merging $0.6+0.9 \msol$ WD binary, 360 s (6 initial orbital periods) after the $0.6 \msol$ WD has been disrupted \citep{dan11}.}
	\label{fig:wvsm}
\end{figure}

As a fiducial example, we show results from an SPH simulation of a merging $0.6+0.9 \msol$ double WD binary performed by \cite{dan11}, which is qualitatively representative of other SPH calculations of unequal mass mergers.  The WDs are assumed to be rotating synchronously \citep{fl11b}.  The solid line in Figure \ref{fig:wvsm} shows the angular velocity, $\Omega$, of material within $10^8$ cm of the equatorial midplane, taken with respect to the center of the $0.9 \msol$ WD, vs. total cylindrically enclosed mass 360 s after the $0.6 \msol $ WD has been disrupted.  Also shown is the Keplerian angular velocity, $\Omega_K = (G M_{\rm enc} / r_{\rm cyl}^3)^{1/2}$ (\emph{dashed line}), where $M_{\rm enc}$ is the mass enclosed within a sphere of radius $r_{\rm cyl}$.  The dotted line shows their ratio.  Following the merger, centrifugal forces are the dominant support against gravity for the outer $0.4 \msol$ of material.  Note that the outer $0.1 \msol$ exists as a spiral tidal tail and thus does not appear fully Keplerian.

\begin{figure}
	\plotone{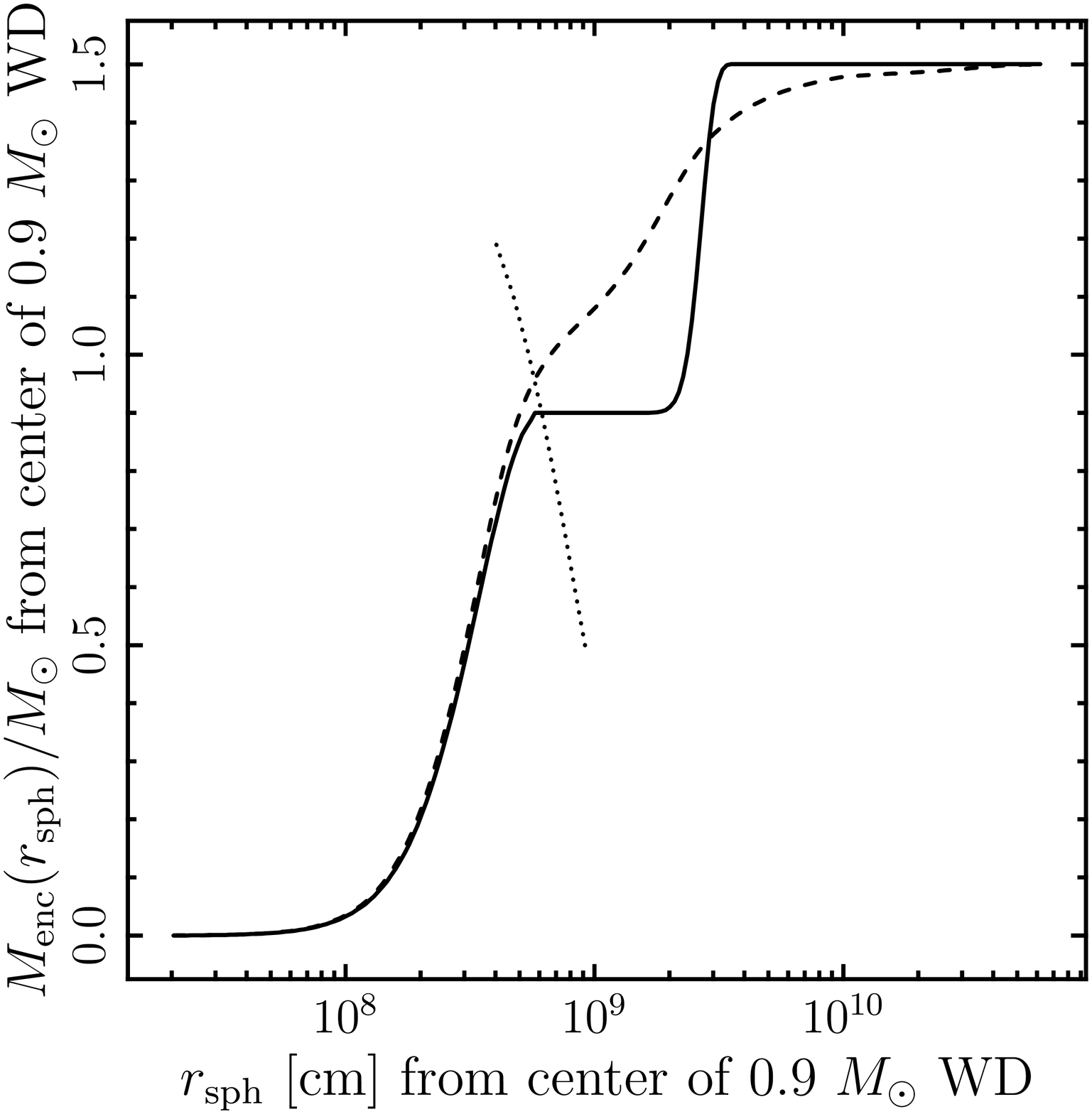}
	\caption{Spherically enclosed mass vs. spherical radius from the center of the $0.9 \msol$ WD in \cite{dan11}'s SPH simulation.  The solid line shows results from the beginning of the simulation, just before mass transfer begins.  The dashed line shows results 360 s after the $0.6 \msol$ WD has been disrupted.  Also shown is the zero-temperature mass-radius relation for C/O WDs (\emph{dotted line}).}
	\label{fig:mvsr}
\end{figure}

Figure \ref{fig:mvsr} shows the enclosed mass vs. spherical radius from the center of the $0.9 \msol$ WD.  The solid line shows the results from the beginning of the simulation, just before mass transfer begins.  Mass transfer continues at increasing levels for 1840 s (30 initial orbital periods) until the less massive WD is disrupted.  The dashed line shows the enclosed mass vs. radius an additional 360 s after the $0.6 \msol$ WD has become unbound.

The enclosed mass vs. radius for most of the inner $0.9 \msol$ of the remnant remains largely undisturbed because most of the tidally disrupted WD remains supported against gravity by rotation, with a smaller contribution from thermal pressure.  However, a small amount of the disrupted WD does exert some pressure on the $0.9 \msol$ core.  The dotted line shows the zero-temperature mass-radius relation for C/O WDs, which intersects the ``before'' (\emph{solid}) and ``after'' (\emph{dashed}) lines at masses that differ by $0.05 \msol$.

Some recent studies have focused on the dynamical mergers of binaries with equal or nearly equal mass components.  In contrast to the SPH results we use throughout this paper, these studies have found that both components, instead of just one, are tidally disrupted \citep{lig09,vcj10} or that the resulting merger is more violent and reaches higher temperatures than in the unequal mass case \citep{pakm10,pakm11}.  However, it is possible that these differences are artifacts of the approximate initial conditions these particular equal mass studies use.  Namely, these simulations begin with the binary components close enough that strong mass transfer immediately sets in once the calculation is begun.  In contrast, \cite{dan11} emphasize the importance of beginning such simulations at larger orbital separations and instead find tidal disruption at a much larger radius with correspondingly less violence.  Given the current uncertainties, we proceed under the assumption that only one WD is tidally disrupted during the merger.  We further assume that the extreme conditions necessary for a detonation are not met during the merger \citep{pakm10,pakm11,guil10}.


\section{Viscous evolution: $10^4-10^8$ \lowercase{s}}
\label{sec:viscevol}

Once the merger remnant has evolved for several dynamical timescales, it will reach quasi-hydrostatic equilibrium.  As Figure \ref{fig:wvsm} shows, considerable differential rotation exists in the disrupted WD's material, more than half of which is nearly Keplerian.  This shearing in the presence of viscous mechanisms naturally leads to outward angular momentum transport.  Quantitative results of this viscous evolution require a multi-dimensional magnetohydrodynamic calculation; such two- and three-dimensional simulations with ZEUS \citep{sn92a} are currently underway.  However, some general points can be made.

Application of an $\alpha$-viscosity prescription \citep{ss73} in the geometrically thick near-Keplerian material, e.g. due to the magnetorotational instability \citep{bh91,hb91,balb03}, yields a viscous timescale of
\be
	t_{\rm visc,\alpha} &\sim& \frac{1}{\alpha} \lp \frac{r_{\rm cyl}}{h} \rp^2 \sqrt{ \frac{r_{\rm cyl}^3}{G M_{\rm enc}} } \nonumber \\
	&\sim& 3\E{4} {\rm \ s} \lp \frac{10^{-2}}{\alpha} \rp \lp \frac{0.1}{h/r_{\rm cyl}} \rp^2  \nonumber \\
	&& \times \lp \frac{r_{\rm cyl}}{10^9 {\rm \ cm}} \rp^{3/2}  \lp \frac{1 \msol }{M_{\rm enc}} \rp^{1/2} ,
	\label{eqn:tvisca}
\ee
where $r_{\rm cyl}$ is the cylindrical radius, and a typical pressure scale height in this rotating material is $h \sim 0.1 r_{\rm cyl}$.  Magnetic stresses may also yield efficient angular momentum transport in the non-rotationally-supported material that has been spun up during the merger (e.g., the material between $0.9-1.0 \msol$ in Fig. \ref{fig:wvsm}).  While the magnetorotational instability cannot operate in this regime where $d \Omega / dr > 0$, stresses due to the winding up of a seed magnetic field by the Tayler-Spruit dynamo may be very effective \citep{tayl73,spru99,spru02,piro08}.  Since this material is stably stratified by its thermal profile, the appropriate effective diffusivity is \cite{spru02}'s case 1,
\be
	\nu_{\rm TS} &=& r_{\rm cyl}^2 \Omega \lp \frac{\Omega}{N} \rp^{1/2} \lp \frac{K}{r_{\rm cyl}^2 N} \rp^{1/2} \nonumber \\
	&=& 3\E{10} \ \frac{\rm cm^2}{\rm s} \lp \frac{r_{\rm cyl}}{3\E{8} {\rm \ cm}} \rp \lp \frac{\Omega}{0.3 {\rm \ s^{-1}}} \rp^{3/2} \nonumber \\
	&& \times \lp \frac{K}{3\E{5} {\rm \ cm^2 / s} } \rp^{1/2} \lp \frac{1 {\rm \ s^{-1}}}{N} \rp .
	\label{eqn:nuTS}
\ee
The square of the Brunt-V\"{a}is\"{a}l\"{a} frequency in the absence of compositional gradients is
\be
	N^2 = \frac{g}{h} \frac{\chi_T}{\chi_\rho} \lp \nabla_{\rm ad} - \frac{d \ln T}{d \ln P} \rp ,
\ee
where the gravitational acceleration is $g$, the logarithmic derivative of the pressure is $\chi_x \equiv \partial \ln P / \partial \ln x$ with the other intrinsic variables held constant, and the adiabatic temperature gradient is $\nabla_{\rm ad} \equiv \partial \ln T / \partial \ln P$ at constant entropy.  The thermal diffusivity in equation (\ref{eqn:nuTS}) is given by
\be
	K &=& \frac{ 16 \sigma T^3}{3 \kappa \rho^2 c_P} \nonumber \\
	&= & 3\E{5} \  \frac{\rm cm^2}{\rm s} \lp \frac{T}{10^8 {\rm \ K} } \rp^3 \lp \frac{ 10^5 {\rm \ g/cm^3 } }{ \rho } \rp^2 \nonumber \\
	&& \times \lp \frac{ 0.01 {\rm \ cm^2 /g } }{ \kappa } \rp \lp \frac{ 1.5 k / 14 m_p}{ c_P } \rp ,
\ee
where the opacity is $\kappa$, and the specific heat at constant pressure, $c_P$, is due primarily to ions in this mostly degenerate material.  This yields a viscous timescale of
\be
	t_{\rm visc,TS} &\sim& \frac{ r_{\rm cyl}^2 \Omega }{ \nu_{\rm TS} q } \nonumber \\
	& \sim & 5\E{7} {\rm \ s} \lp \frac{r_{\rm cyl}}{3\E{8} {\rm \ cm} } \rp^2 \lp \frac{\Omega}{0.3 {\rm \ s^{-1}}} \rp \nonumber \\
	&& \times \lp \frac{3\E{10} {\rm \ cm^2 / s}}{\nu_{\rm TS}} \rp \lp \frac{0.02 {\rm \ s^{-1} }}{q} \rp ,
\ee
where the shear rate in these layers is $q = d\Omega / d \ln r_{\rm cyl}$.

As long as $\alpha \ll 1$, both the $\alpha$-viscosity and Tayler-Spruit dynamo processes yield viscous timescales that are $ 10^4-10^8$ s, so that our assumption that the merger remnant enters into hydrostatic equilibrium prior to its viscous evolution is well-founded.  Furthermore, as we show in the next section, the viscous timescale is orders of magnitude shorter than the thermal timescale.  Thus, the viscous evolution occurs with no significant heat loss.  Because the material is optically thick and neutrino cooling is not significant on such short timescales, the heat generated dynamically during the merger and through viscous shear heating will remain trapped in the envelope until it can diffuse out on the much longer thermal timescale.

As angular momentum is transported outwards to larger radii, the merger remnant evolves towards an equilibrium state of shear-free solid-body rotation.  During the progression towards this equilibrium, the local specific heating rate due to shear, $\epsilon_{\rm visc} = \nu q^2$, both increases the specific internal energy, $u$, and does expansion work by spreading material to larger radii.  While the precise end state of the remnant is a function of the viscous mechanism and the multi-dimensional time-dependent evolution, local conservation of energy implies that the originally near-Keplerian material is converted to nearly virial material with total specific energy $ u - G M_{\rm enc}/r \simeq  \lp r_{\rm cyl} \Omega_0 \rp^2/2$, where $\Omega_0$ is the initial angular velocity after the dynamical phase of evolution.  In effect, during the course of viscous evolution, the centrifugally-supported material is converted into a hot envelope surrounding the degenerate core.  By the time significant thermal transport begins, most of the less massive WD has been converted into a thermally-supported and extended envelope with negligible rotation.

As an example, we present a one-dimensional $\alpha$-viscosity hydrodynamic simulation of the outer $0.6 \msol$ of \cite{dan11}'s $0.6+0.9 \msol$ merger.  Details of this viscous calculation are presented in the Appendix.  The simulation begins 360 s after the $0.6 \msol$ WD has been disrupted.  We assume the inner $0.9 \msol$ remains unchanged during the viscous evolution.  We evolve the axisymmetric height-integrated Navier-Stokes equations under the assumption that the only entropy evolution term is due to shear heating because of the inefficiency of radiative cooling.  We follow the cylindrical radius, $r_{\rm cyl}$, radial and angular velocities, $v_r$ and $\Omega$, height-integrated surface density, $\Sigma= \int_{-\infty}^\infty \rho dz$, and height-integrated pressure multiplied by a factor of the adiabatic exponent $\gamma$, $\Pi \equiv \gamma \int_{-\infty}^\infty P dz$, for 200 equal mass Lagrangian cylindrical shells.

We assume $\alpha = 10^{-2}$ throughout the material, even though the physical source of viscosity changes from the inner regions to the outer.  While this could alter the precise values at the end of the simulation, the qualitative results will be similar as long as the viscous timescale is intermediate to the dynamical and thermal timescales.  A $\gamma=5/3$ equation of state is assumed to hold throughout the disk, which becomes questionable when radiation pressure becomes important at later times.  However, simulations with smaller values of $\gamma$ have qualitatively similar outcomes.

\begin{figure}
	\plotone{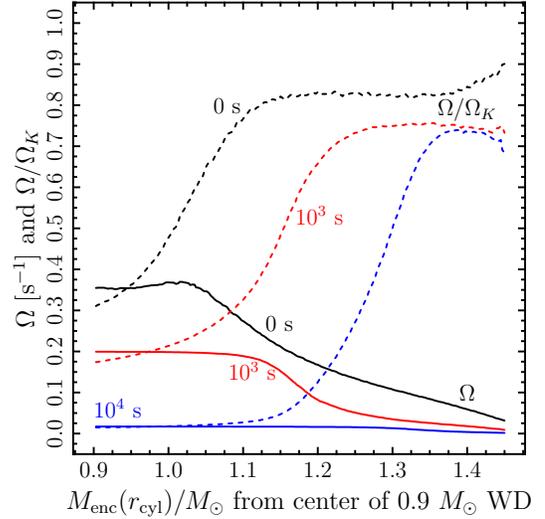}
	\caption{Measures of the angular velocity vs. cylindrically enclosed mass at different times during the 1D viscous simulation.  Solid lines show the angular velocity, and dotted lines show the ratio of $\Omega/\Omega_K$ at 0, $10^3$, and $10^4$ s after the beginning of the simulation.}
	\label{fig:wvsmvisc}
\end{figure}

Figure \ref{fig:wvsmvisc} shows the resulting angular velocity vs. the cylindrically enclosed mass $0$, $10^3$, and $10^4$ s after the beginning of our simulation (\emph{solid lines}).  As expected from equation (\ref{eqn:tvisca}), the rotational profile evolves towards a shear-free configuration within a viscous timescale $\sim 10^4$ s.  Also plotted are the ratios of $\Omega/\Omega_K$ at these different times (\emph{dotted lines}), showing that the material originally from the tidally disrupted WD evolves from being mostly rotationally-supported to mostly pressure-supported.

\begin{figure}
	\plotone{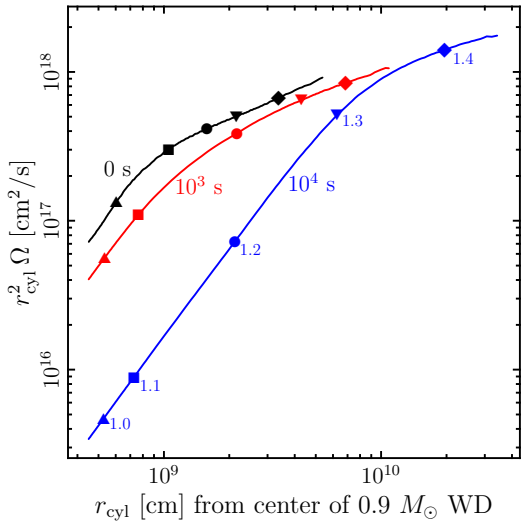}
	\caption{Specific angular momentum vs. cylindrically enclosed radius at the same snapshots in time as in Figure \ref{fig:wvsmvisc}.  Markers denote cylindrically enclosed masses of $1.0-1.4 \msol$ separated by $0.1 \msol$, as labeled for the $10^4$ s snapshot.}
	\label{fig:jvsrvisc}
\end{figure}

\begin{figure}
	\plotone{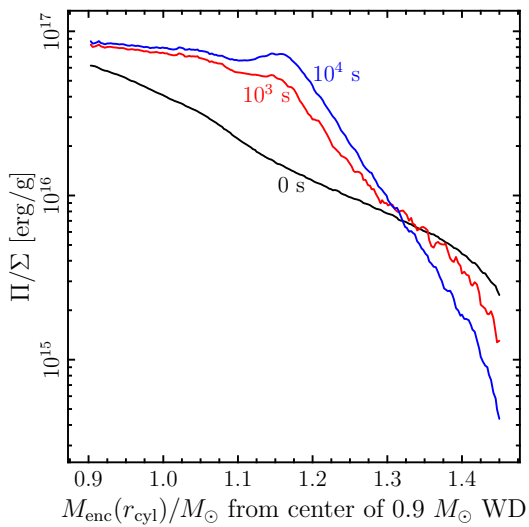}
	\caption{A proxy for the equatorial specific internal energy, $\Pi / \Sigma$, vs. cylindrically enclosed mass at the same snapshots in time as in Figure \ref{fig:wvsmvisc}.  The actual equatorial internal energy cannot be precisely determined from the simulation because we only follow height-integrated quantities.}
	\label{fig:uvsmvisc}
\end{figure}

Figure \ref{fig:jvsrvisc} shows the specific angular momentum profile vs. cylindrical radius at the same snapshots in time as in Figure \ref{fig:wvsmvisc}.  Markers denote cylindrically enclosed masses as labeled.  It is clear that as angular momentum is transferred outwards, the outer regions $\gtrsim 1.2 \msol$ evolve by spreading.  Figure \ref{fig:uvsmvisc} shows the ratio of $\Pi / \Sigma$ vs. cylindrically enclosed mass at the same snapshots in time as in Figure \ref{fig:wvsmvisc}.  This ratio is a proxy for the specific internal energy at the equator; calculation of the actual equatorial specific energy requires knowledge of the remnant's vertical structure, which is not followed in the height-integrated approximation.  As Figure \ref{fig:uvsmvisc} qualitatively demonstrates, the internal energy between $1.0-1.3 \msol$ increases significantly as rotational energy is converted into heat.  Because of this thermal pressure support, material within $1.3 \msol$ does not move inwards significantly and compress the core.  On the other hand, material outside of $1.3 \msol$ spreads outwards considerably, so its internal energy drops via adiabatic expansion.

While the main qualitative results of this simulation should be robust $-$ the outwards transport of angular momentum, the conversion of rotational support into thermal pressure support, and the spreading of the outer regions $-$ the quantitative results are only approximate.  This is due to the height-integrated approximation, the confinement of mass to cylindrical shells, and the hard inner surface boundary condition.  For example, as angular momentum is removed from the inner regions of the simulation, some mass should flow towards the poles of the degenerate core at $r_{\rm cyl}=0$.  This approach to spherical symmetry, and other multi-dimensional results, await a future simulation with the necessary physics (J. Schwab, et al., in preparation).


\section{Thermal evolution: $10^3- 10^4$ \lowercase{yr}}
\label{sec:thermevol}

After the merger remnant has viscously evolved to a slowly rotating shear-free equilibrium, its subsequent evolution is determined by the heat generated previously during the dynamical and viscous phases of the merger.  If the post-viscous temperature peak is sufficiently high ($\gtrsim 6\E{8}$ K at a density of $5\E{5} \cgsd$), a convective C-burning shell will develop immediately following the viscous evolution on a timescale of $\sim 10^6 $ s, prior to any significant thermal transport.  While we are unable to extract precise temperatures from our one-dimensional viscous simulation, our preliminary multi-dimensional results suggest that this is indeed marginally the case for our fiducial $0.6+0.9 \msol$ example.  Such off-center convective burning converts the degenerate C/O core into a C-burning star from the outside in, eventually yielding a degenerate O/Ne WD \citep{ni85,twt94,sn98}.  During the $\sim 10^4$ yr it takes for the burning layer to diffuse into the center, the shell-burning merger remnant radiates a small fraction of the nuclear luminosity and appears as a near-Eddington source \citep{sn98}.

For binary components of lower masses and thus lower gravitational potentials, the peak temperature following the viscous evolution will be lower, and convective C-burning will not immediately take place.  In these cases, the further evolution will be determined by the transport of heat through the hot envelope.  Since the envelope consists of material that was initially only halfway bound to the undisrupted WD, the conversion of its gravitational energy to heat and expansion work implies that the material is at least somewhat virialized, with a pressure scale height at the envelope's base that is a non-negligible fraction of the degenerate core's radius: $h \sim R_{\rm core}$.

A simple order of magnitude estimate implies that the ratio of radiation to gas pressure in a virialized envelope is
\be
	\frac{P_{\rm rad}}{P_{\rm gas}}\sim 0.8 \mu^4 \lp \frac{ M_{\rm core} }{M_\odot }\rp^3 \lp \frac{ M_\odot}{M_{\rm env} } \rp ,
\ee
where $\mu$ is the mean molecular weight.  For double C/O WD mergers, this implies that radiation pressure is at least comparable to ideal gas pressure in the envelope.\footnote{However, this is not the case for lower mass cores and lower mean molecular weights (e.g., a $1 \msol$ star on the RGB or low mass double He WD merger remnants).}  Thus, the remnant radiates at a significant fraction of the Eddington luminosity while entropy is redistributed within the envelope.  This occurs on a thermal timescale that is of order
\be
	\label{eqn:ttherm}
	t_{\rm therm} &\sim& \frac{GM_{\rm core} M_{\rm env} /R_{\rm core} }{4\pi G ( M_{\rm core} + M_{\rm env} ) c / \kappa} \nonumber \\
	&\sim& 10^4 {\rm \ yr} \lp \frac{\kappa}{0.2 {\rm \ cm^2/g}} \rp \lp \frac{10^9 {\rm \ cm}}{R_{\rm core}} \rp 
\ee
for core and envelope masses $M_{\rm core} = 0.8 \msol$ and $M_{\rm env} = 0.6 \msol$.  This timescale is equivalent to the Kelvin-Helmholtz contraction timescale for the material at the base of the envelope near $R_{\rm core}$, which transfers its entropy into regions further out in the envelope in the process of contracting onto the core.

To explore this phase of thermal transport in more detail, we utilize the spherically symmetric stellar evolution code MESA \citep{paxt11}\footnote{http://mesa.sourceforge.net/ (version 3635)}.  Because our height-integrated cylindrically symmetric viscous evolution (see \S \ref{sec:viscevol}) cannot produce spherically symmetric results, we begin our thermal evolution with rough, but qualitatively correct, initial conditions: a degenerate $0.8 \msol$ core at an initial temperature of $10^8$ K surrounded by a high entropy $0.6 \msol$ envelope that does not exert significant pressure on the core.  A primary of $0.8\msol$ was chosen for its relatively lower gravitational potential, so that the maximum temperature in the post-viscous merger remnant would be lower than in the $0.9+0.6 \msol$ case.  As previously mentioned, multi-dimensional viscous simulations are underway with ZEUS (J. Schwab, et al., in preparation); future work will use these results as initial conditions for a more quantitatively correct thermal evolution.

Given our poor understanding of mass loss in radiation-dominated H- and He-deficient envelopes, we turn off the options for mass loss in MESA.  Additionally, the convective mixing length parameter is increased by a factor of 10 in the outer 0.1\% of the remnant's mass in order to avoid severely pathological density profiles near the photosphere of the hot envelope, which otherwise force the code to take tiny timesteps.  We will address these simplifications in more detail in \S \ref{sec:conc}.

\begin{figure}
	\plotone{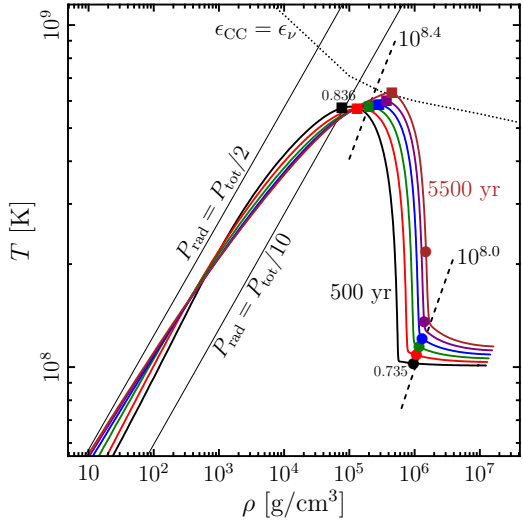}
	\caption{Snapshots of the $T-\rho$ profile during the thermal evolution of a $0.6+0.8 \msol$ merger remnant (\emph{thick solid lines}).  Profiles are equally spaced by 1000 yr and begin $500$ yr after the end of the viscous evolution.  Convective C-burning begins at $5790$ yr at a mass coordinate of $0.836 \msol$, which is represented by a square for each snapshot.  The mass coordinate $0.735 \msol$ is represented by bullets.  The dashed lines marks contours of constant specific entropy $10^{8.0}$ and $10^{8.4}$ erg g$^{-1}$ K$^{-1}$ as labeled.  The thin solid lines denote $P_{\rm rad} = P_{\rm tot}/10$ and $P_{\rm tot}/2$.  The dotted line marks the contour for which heating from C-burning equals neutrino cooling.}
	\label{fig:tvsrho}
\end{figure}

\begin{figure}
	\plotone{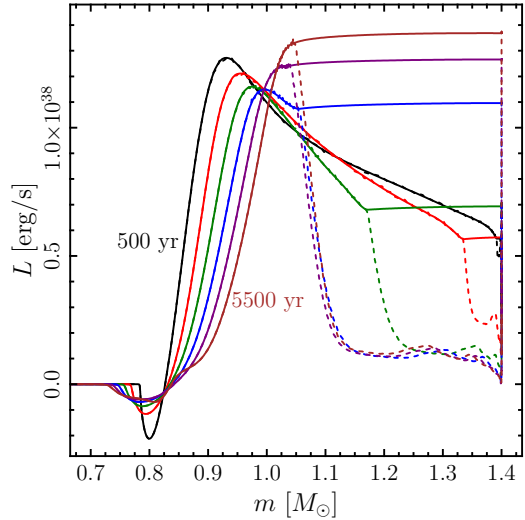}
	\caption{The total (\emph{solid lines}) and radiative (\emph{dashed lines}) luminosities vs. mass at the same snapshots in time as in Figure \ref{fig:tvsrho}.  The Eddington luminosity is $3.5\E{38}$ erg s$^{-1}$ for a remnant mass of $1.4 \msol$ and $\kappa=0.2$ cm$^2$ g$^{-1}$.}
	\label{fig:lvsm}
\end{figure}

\begin{figure}
	\plotone{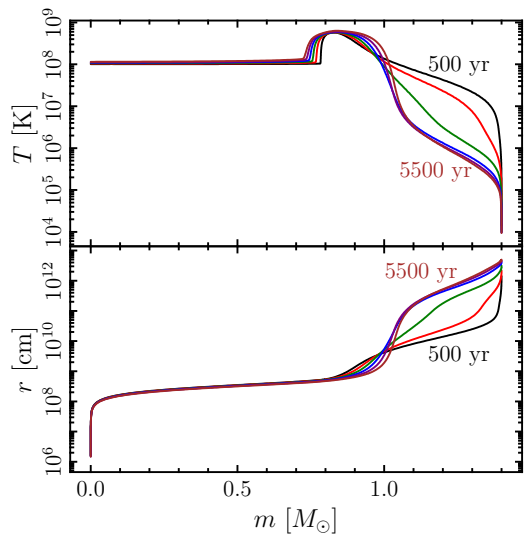}
	\caption{Temperature and radius vs. mass profiles at the same snapshots in time as in Figure \ref{fig:tvsrho}.}
	\label{fig:trvsm}
\end{figure}

Figure \ref{fig:tvsrho} shows $T-\rho$ profiles (\emph{thick solid lines}) at six snapshots in time equally spaced by 1000 yr, starting 500 yr after the thermal evolution begins.  The dotted line marks the contour for which heating from C-burning matches neutrino cooling.  The thin lines show conditions for which the radiation pressure, $P_{\rm rad}$, equals one-half and one-tenth of the total pressure, $P_{\rm tot}$.  Bullets correspond to the trajectory of a fluid element at a mass coordinate of $0.735 \msol$; squares correspond to a mass coordinate of $0.836 \msol$.  The dashed lines demarcate contours of constant specific entropy at $10^{8.0}$ and $ 10^{8.4}$ erg g$^{-1}$ K$^{-1}$.  As the squares show, fluid elements near the temperature peak lose entropy during the thermal evolution.  A small amount of this entropy diffuses inwards into the degenerate core, reaching a mass coordinate of $0.735 \msol$ after $6000$ yr.  However, since the diffusion timescale to the center of the degenerate core is $ \sim 10^6$ yr, the majority of the entropy from the inner regions is instead transferred outwards through the envelope.  As this material near the temperature peak loses entropy, it undergoes a Kelvin-Helmholtz contraction, falling from the hot extended envelope to the cold degenerate core and compressing to higher densities.  This compression more than compensates for the loss of entropy and causes these fluid elements to increase in temperature.

After 5790 yr have passed, the peak temperature reaches $ 6.3\E{8}$ K at a density of $4\E{5} \cgsd$, which is high enough for nuclear burning to overcome neutrino cooling and radiative diffusion, resulting in the onset of off-center convective C-burning at a mass coordinate of $0.836 \msol$.  Thus, even if convective shell-burning is avoided immediately following the viscous evolution and prior to significant thermal transport, it eventually sets in due to the cooling and compression of the base of the hot envelope.

The key property of the post-viscous evolution that determines the time at which off-center convective burning begins is the conditions at the temperature peak.  We have chosen a somewhat arbitrary starting point for our thermal evolution calculation, with an initial temperature maximum of $6.0\E{8}$ K at $10^5 \cgsd$.  If instead the post-viscous temperature peak were located at a higher density of $3\E{5} \cgsd$, only $ \sim 1000$ yr would pass before convection begins.

During the thermal evolution, the envelope approaches a constant luminosity profile at roughly one-third the Eddington limit, which is $3.5\E{38}$ erg s$^{-1}$ for a remnant mass of $1.4 \msol$ and $\kappa=0.2$ cm$^2$ g$^{-1}$.  In the interior of the envelope, most of this flux is carried by convection, so that the radiative luminosity profile is not constant (see the dashed lines in Fig. \ref{fig:lvsm}).  As entropy is redistributed throughout the envelope to yield a roughly constant total luminosity, shown as solid lines in Figure \ref{fig:lvsm}, the envelope spreads outwards significantly into a giant configuration until the majority of the envelope mass is $ >10^{11}$ cm.  Mass-radius profiles are shown in the bottom panel of Figure \ref{fig:trvsm}.  The top panel of Figure \ref{fig:trvsm} shows that temperatures throughout the extended regions of the envelope are $ \lesssim 10^6$ K due to this expansion.

Given the photospheric radius of $10^{12}-10^{13}$ cm and the near-Eddington luminosity, the effective temperature during this cooling phase ranges from $5000-30000$ K.  However, we again emphasize that our artificial enhancement of the convective mixing length near the photosphere and the restriction on mass loss may change these results.  While the merger remnant will radiate near the Eddington limit for $\sim 10^4$ yr, predictions of its temperature and atmospheric properties will likely change with a more physical treatment of mass loss and convection.  Future work on the thermal evolution of merger remnants will better quantify the appearance of these systems during the cooling and shell-burning phases.


\section{Comparison to earlier work}
\label{sec:comp}

Two previous studies have examined the viscous and thermal evolution of the merger remnant using appropriate initial conditions set by SPH simulations of double WD mergers.  \cite{ypr07} map the output of their SPH simulations to a one-dimensional stellar evolution code that includes the effects of rotation \citep{yl04a}.  Their initial remnant consists of the material primarily supported by pressure, which includes the material that had been shock-heated during the merger.  The material primarily supported by rotation is assumed to reside in a Keplerian disk from which matter is accreted at a prescribed constant rate.  Their viscous evolution, which occurs concurrently with the thermal evolution, includes the effects of hydrodynamic viscous transport processes but assumes that magnetic stresses are negligible.  Furthermore, the specific angular momentum of each shell is assumed to decrease on a parametrized timescale without liberating kinetic energy.

Several important differences distinguish our model from \cite{ypr07}'s, although some aspects are similar.  The most significant physical difference is their restriction to hydrodynamic angular momentum transport processes.  If magnetic stresses are assumed to be negligible, the timescale for angular momentum transport becomes comparable to the thermal timescale.  As material from their rotating hot envelope cools and moves inwards, rotational support plays an increasingly important role and slows the contraction to the relatively slow angular momentum transport timescale, allowing more time for neutrino cooling and inwards thermal diffusion.  Another important difference is their assumption that the material primarily supported by rotation following the merger resides in a Keplerian disk during the remnant evolution.  We show that, instead, most of this material is converted into a high-entropy extended envelope during a phase of viscous evolution that occurs prior to the thermal evolution (see Fig. \ref{fig:wvsmvisc}).

Several of \cite{ypr07}'s models assume accretion from the disk is negligible and that angular momentum loss from the total remnant occurs rapidly, so that angular momentum transport within the remnant is unimportant (e.g., sequences Sa2, Aa2, and Ba2).\footnote{Their test model Ta1 has no initial rotation nor accretion and would be very similar to our model, but the material at Ta1's initial temperature peak is degenerate, with $k T / E_F= 0.03$, where $E_F$ is the Fermi energy.  Our model begins much more non-degenerately because of previous viscous heating, with $kT/E_F=1$ at the temperature peak, and thus it experiences a much stronger compressional evolution.}  Our model resembles these calculations if the total binary mass is assumed to be equal to their central remnant mass and if the heat generated by viscous dissipation is ignored.  Due to these similarities, \cite{ypr07} also find that off-center C-ignition occurs in these models because of the relatively rapid Kelvin-Helmholtz contraction of the hot envelope surrounding the degenerate core.

\cite{vcj10} consider the viscous evolution of mergers of equal-mass WDs in which both WDs are tidally disrupted, utilizing data from \cite{lig09}'s SPH simulations.  The resulting remnant has a temperature profile that peaks at the center, unlike remnants in which only one WD is disrupted, which have a temperature peak in material at the edge of the degenerate core.  As in our work, their model assumes that magnetic stresses transport angular momentum in the outer material very rapidly.

However, \cite{vcj10} neglect the conversion of the rotational energy into heat, so that this material is rapidly transformed into degenerate material that exerts pressure on the core, yielding explosive C-burning at the center.  While we agree that rapid viscous evolution occurs without significant heat transport between fluid elements, local conservation of energy demands that viscous dissipation converts the nearly Keplerian rotational energy into an equivalent amount of heat and expansion work.  The resulting thermal pressure support prevents the outer material from compressing the central core until it undergoes Kelvin-Helmholtz contraction to rid itself of this excess heat.  Furthermore, as mentioned in \S \ref{sec:dynevol}, it is unclear that SPH simulations with accurate initial conditions allow for the tidal disruption of both WDs, even if they have equal masses.  Because mass transfer in more accurate simulations begins at a larger orbital separation and persists for multiple orbital periods prior to the merger, there is opportunity for symmetry breaking and the possibility that only one of the WDs is disrupted.  Ongoing work will help to explore these outcomes in detail (M. Dan, et al., in preparation).


\section{Conclusions}
\label{sec:conc}

We have examined the post-merger evolution of unequal mass double C/O WD systems taking into account the magnetic stresses that redistribute angular momentum rapidly compared to the thermal evolution of the remnant.   This divides the merger evolution into a hierarchy of timescales: a rapid dynamical phase ($10^2-10^3$ s) in which the less massive WD is tidally disrupted; an intermediate viscous evolution ($10^4-10^8$ s) during which the rotating envelope spreads and viscously heats without cooling; and, if convective C-burning does not set in first, a secular thermal phase ($10^3- 10^4$ yr) in which the hot extended envelope radiates and contracts.  This model is in contrast to earlier work that assumed the tidally disrupted WD is transformed into a Keplerian accretion disk from which the more massive WD accretes at a constant rate \citep{ni85,sn98,sn04,pier03b,pier03a}.

However, while the sequence of events is different, the outcome is qualitatively similar to previous work.  In our models that do not immediately begin convective burning following the viscous evolution, thermal transport within and outwards from the envelope causes the base to be compressed, while in accretion models, accreting material is responsible for this compression.  In both cases, however the rate of compression is set by the Eddington luminosity limit.  Because of this relatively rapid compression and the prior dynamical and viscous heating, off-center convective C-burning is initiated and no SN Ia follows.  Instead, the more likely outcome is the formation of a high-mass O/Ne WD and possibly the collapse of the WD to a neutron star, as argued in previous calculations.   Although this is our preliminary conclusion, the physical model presented here differs significantly from that in previous work, so there are a number of unresolved issues that need to be explored in more detail before the fate of WD mergers can be definitively established.

In particular, the actual thermal evolution of the merger remnant may be strongly influenced by mass loss, which we have neglected in this study.  Dust-driven, line-driven, and, given the importance of radiation pressure in the Eddington-limited envelope, continuum radiation-pressure-driven winds may play a significant role.  In addition to determining the thermal evolution of the inner layers, these winds may also influence the spectrum of a subsequent SN Ia.  The escape velocity for material at the outer edge of the remnant ($\sim10^{13}$ cm) during the thermal evolution is  $\sim 60$ km s$^{-1}$.  If winds and/or the mass ejected during the coalescence and the viscous evolution propagate outwards for $10^4$ yr, material will fill the surrounding volume out to $\sim 2\E{18}$ cm at the time of an explosion.  Such an outflow could possibly explain the variable (e.g., \citealt{simo09}) or preferentially blue-shifted \citep{ster11} Na I D absorption lines seen in some SNe Ia, although further detailed study of the mass loss mechanism is necessary to make quantitative predictions.

Due to this neglect of mass loss, the eventual fate of the merger remnant remains uncertain.  However, if the remnant collapses to a neutron star or yields a SN Ia, the resulting light curves will be affected by the presence of the extended envelope if it persists until the explosion.  As ejecta from either of these events runs into $\sim 0.1 \msol$ of material at radii of $\sim 10^{13}$ cm, the resulting shock efficiently converts kinetic energy into thermal energy that will then be radiated on a photon diffusion timescale of several days.  This makes a merger's collapse to a neutron star easier to observe; instead of reaching maximum light at $\simeq 1$ d, such ``enshrouded'' explosions would peak at several days and decay from peak more slowly as well \citep{darb10,mpq09b}.  However, in the case of SNe Ia, such a large amount of observable shock-deposited energy has been ruled out in at least one case by recent early-time measurements of SN 2011fe \citep{nuge11,bloo12}.  If SNe Ia arise from double WD mergers, their extended envelopes must not persist until the time of explosion.

Future work is required to explore these outcomes, as well as to correctly calculate the envelope's structure near the photosphere.  Our artificial enhancement of the convective mixing length near the surface allows us to overcome pathological density profiles, but it prevents us from reliably calculating the effective temperature of the remnant.  Even so, we can make several robust predictions from our model.  Double C/O WD merger remnants will radiate near the Eddington limit for $ \gtrsim 10^4$ yr with bolometric luminosities $\sim 10^{38}$ erg s$^{-1}$.  A definitive statement regarding their spectra and colors during their thermal evolution cannot be made without better determination of the mass loss rate.  If outflowing dust does indeed play a significant role, they may be infrared-bright and might be identified in a sample such as that of \cite{tiss11}.  Given the double WD merger rate of a few per $1000$ yr \citep{nele01a,rbf09}, a Milky Way-type galaxy should harbor $ \gtrsim $ a few dozen of these bright H- and He-deficient sources at any time.

In this work, we have focused primarily on the merger of two C/O WDs.  However, the same physical picture will also describe mergers for which one or both of the components is a He WD.  Double He WD mergers will undergo a very similar evolution as we have described for C/O WD mergers, with appropriately adjusted energy and timescales: a He-burning shell will ignite off-center and diffuse inwards, possibly with episodic flashes due to the inefficiency of neutrino cooling \citep{iben90,sn98}.  Once the burning shell reaches the center, the merger remnant appears as a core He-burning sdB or sdO star and eventually cools to become a C/O WD.  The merger of a He and a C/O WD should yield a He shell burning R CrB star \citep{webb84,iben90,clay96}, although the final He shell flash of cooling post-AGB stars has also been proposed as an R CrB progenitor \citep{ity96}.  Simulations of these mergers within the framework of our updated physical model will allow us to connect theoretical predictions with observational constraints \citep{aspl97,clay07,clay11,jks11} and help to differentiate between these progenitor scenarios.


\acknowledgments

We thank the anonymous referee for comments that helped to guide and clarify our work.  We are very grateful to M. Dan and S. Rosswog for access to their simulation data.  We thank them, R. Foley, W. Haxton, B. Metzger, T. Piro, P. Podsiadlowski, J. Schwab, S. Sim, and M. van Kerkwijk for helpful discussions.  We also thank B. Paxton and the MESA Council for providing and tirelessly developing MESA.  This work was supported by the National Science Foundation under grants PHY 05-51164 and AST-11-09174, the U.S. Department of Energy under grant DE-SC00046548, and the Theoretical Astrophysics Center at UC Berkeley.   KJS is supported by NASA through the Einstein Postdoctoral Fellowship awarded by the Chandra X-ray Center, which is operated by the Smithsonian Astrophysical Observatory for NASA under contract NAS8-03060.  EQ is supported in part by the David and Lucile Packard Foundation.  DK's work is supported by the Director, Office of Energy Research, Office of High Energy and Nuclear Physics, Divisions of Nuclear Physics, of the U.S. DOE under grant DE-AC02-05CH11231 and by the DOE SciDAC Program (DE-FC02-06ER41438).


\

\

\

\section*{Appendix: Viscous simulation details}
\label{sec:app}

In this Appendix, we explain the computational details of our height-integrated viscous simulation whose results are shown in \S \ref{sec:viscevol}.  The approximations outlined there yield the Lagrangian time derivatives:
\be
	\frac{ d\Sigma}{dt} = - \frac{\Sigma}{r_{\rm cyl}} \frac{d \lp r_{\rm cyl} v_r \rp }{d r_{\rm cyl} } ,
\ee

\be
	\frac{ dr_{\rm cyl}}{dt} = v_r ,
\ee

\be
	\frac{dv_r}{dt} = -\frac{2 \pi r_{\rm cyl}}{\gamma } \frac{d \Pi}{d m} + r_{\rm cyl} \lp \Omega^2 - \Omega_K^2 \rp ,
\ee

\be
	\frac{ d \Omega }{d t} \simeq - \frac{ 2 \Omega v_r}{r_{\rm cyl}} + \frac{4 \pi^2 \alpha }{ r_{\rm cyl}^2} \frac{d }{d m } \lp \frac{ \Pi \Sigma r_{\rm cyl}^4  }{\Omega_K}  \frac{d \Omega}{d m}  \rp , {\rm \ and}
\ee

\be
	\frac{d \Pi }{dt} &=& - \frac{ \gamma \Pi v_r}{r_{\rm cyl}}  - 2 \pi r_{\rm cyl} \gamma \Pi \Sigma  \frac{ d v_r }{d m}  \nonumber \\
	&& +  \frac{\gamma (\gamma-1) \alpha \Pi }{\Omega_K} \lp 2 \pi r_{\rm cyl}^2 \Sigma \frac{d \Omega}{d m} \rp^2 ,
\ee
where the cylindrically enclosed mass is $m$.

The height-integrated quantities $\Sigma$ and $\Pi$ are defined at zone centers, while the mass, radius, $v_r$, and $\Omega$ are defined at zone boundaries.  The radius at the inner boundary of the grid, where the enclosed mass is $0.9 \msol$, is assumed to be fixed; i.e., the degenerate core is assumed to be incompressible.  The angular velocity gradient across the inner and outer boundaries is set to zero.  These boundary conditions, while clearly approximations, do conserve the total energy and angular momentum in the grid to better than 1\% during the simulation.

We approximate the radial gravitational acceleration at $(r_{\rm cyl},z)$, where $z$ is the height from the midplane, as the value at the midplane, $f_r(r_{\rm cyl}, z) \simeq -G m / r_{\rm cyl}^2  $.
This is equivalent to the assumption that $\int_{-\infty}^\infty \rho z^2 dz \ll \Sigma r_{\rm cyl}^2 $.  

High-frequency spurious noise in $v_r$ is damped with the standard \cite{vr50} artificial viscosity prescription modified for application to curvilinear coordinates \citep{tw79}.  We define an artificial viscous pressure term as
\be
	\mathbb{Q}_{ij} =  \left\{ \begin{array}{lc}
		\rho l^2  \grad \cdot \vec{v} \left[ \lp \grad \vec{v} \rp_{ij} - \frac{1}{3} \delta_{ij} \grad \cdot \vec{v} \right]  & {\rm \ for \ } \grad \cdot \vec{v} < 0 \\
		0 & {\rm \ for \ } \grad \cdot \vec{v} > 0 \\
	\end{array} \right. ,
	\label{eqn:artifvisc}
\ee
and we add the terms $ - \grad \cdot \mathbb{Q} $ and $- \mathbb{Q} \cdot (\grad \vec{v})$ to the momentum and entropy equations, respectively.  The symmetrized velocity gradient tensor is
\be
	\lp \grad \vec{v} \rp_{ij} \equiv \frac{1}{2} \lp \dd{v_i }{x_j} + \dd{ v_j }{x_i } \rp ,
\ee
and $\delta_{ij}$ is the standard Kronecker delta,.  The parameter $l$ is a smoothing length, chosen to be the minimum zone spacing at each time step, and is constant throughout the grid to preserve $ {\rm Tr} [ \mathbb{Q}]=0$ \citep{sn92a}.  The condition on $\grad \cdot \vec{v}$ in equation (\ref{eqn:artifvisc}) ensures that the artificial viscosity only acts during compression.



\end{document}